\documentclass[%
 aip,
 amsmath,amssymb,
 reprint,%
]{revtex4-1}

\usepackage{graphicx}
\usepackage{dcolumn}
\usepackage{bm}

\usepackage[utf8]{inputenc}
\usepackage[T1]{fontenc}
\usepackage{mathptmx}
\usepackage{etoolbox}

\makeatletter
\def\@email#1#2{%
 \endgroup
 \patchcmd{\titleblock@produce}
  {\frontmatter@RRAPformat}
  {\frontmatter@RRAPformat{\produce@RRAP{*#1\href{mailto:#2}{#2}}}\frontmatter@RRAPformat}
  {}{}
}%
\makeatother
\begin{document}

\preprint{AIP/123-QED}

\title[Temporal Evolution of the Light Emitted by a Thin, Laser-ionized Plasma Source]{Temporal Evolution of the Light Emitted by a Thin, Laser-ionized Plasma Source}
\author{Valentina Lee}
\affiliation{Center for Integrated Plasma Studies, Department of Physics, University of Colorado Boulder, Boulder, Colorado 80309, USA}
\author{Robert Ariniello}
\affiliation{SLAC National Accelerator Laboratory, 2575 Sand Hill Rd, Menlo Park, CA 94025, USA}
\author{Christopher Doss}
\affiliation{Center for Integrated Plasma Studies, Department of Physics, University of Colorado Boulder, Boulder, Colorado 80309, USA}
\author{Kathryn Wolfinger}
\affiliation{RadiaSoft LLC, 1790 38th St Suite 306, Boulder, CO 80301, USA}
\author{Peter Stoltz}
\affiliation{Zap Energy, 5901 23rd Dr W Everett, WA 98203, USA}
\author{Claire Hansel}
\affiliation{Center for Integrated Plasma Studies, Department of Physics, University of Colorado Boulder, Boulder, Colorado 80309, USA}
\author{Spencer Gessner}
\affiliation{SLAC National Accelerator Laboratory, 2575 Sand Hill Rd, Menlo Park, CA 94025, USA}
\author{John Cary}
\affiliation{Center for Integrated Plasma Studies, Department of Physics, University of Colorado Boulder, Boulder, Colorado 80309, USA}
\affiliation{Tech-X, 5621 Arapahoe Avenue Suite A Boulder, CO 80303, USA}
\author{Michael Litos}
\affiliation{Center for Integrated Plasma Studies, Department of Physics, University of Colorado Boulder, Boulder, Colorado 80309, USA}

\date{\today}

\begin{abstract}

We present an experimental and simulation-based investigation of the temporal evolution of light emission from a thin, laser-ionized Helium plasma source. We demonstrate an analytic model to calculate the approximate scaling of the time-integrated, on-axis light emission with the initial plasma density and temperature, supported by the experiment, which enhances the understanding of plasma light measurement for plasma wakefield accelerator (PWFA) plasma sources. Our model simulates the plasma density and temperature using a split-step Fourier code and a particle-in-cell (PIC) code. A fluid simulation is then used to model the plasma and neutral density, and the electron temperature as a function of time and position. We then show the numerical results of the space-and-time-resolved light emission and that collisional excitation is the dominant source of light emission. We validate our model by measuring the light emitted by a laser-ionized plasma using a novel statistical method capable of resolving the nanosecond-scale temporal dynamics of the plasma light using a cost-effective camera with microsecond-scale timing jitter. This method is ideal for deployment in the high radiation environment of a particle accelerator that precludes the use of expensive nanosecond-gated cameras. Our results show that our models can effectively simulate the dynamics of a thin, laser-ionized plasma source and this work is useful to understand the plasma light measurement, which plays an important role in the PWFA.

\end{abstract}

\maketitle

\section{\label{sec:intro}Introduction}
Plasma-based accelerators have demonstrated accelerating gradients two to three orders of magnitude larger than conventional radio-frequency accelerators, making them an enticing alternative for future high-energy particle accelerator applications~\cite{Blumenfeld2007, Litos2014, Litos2016}. In a beam-driven plasma wakefield accelerator (PWFA), an electron drive beam generates a wake as it propagates through a plasma. A second electron beam follows the drive beam at a distance on the order of the plasma skin depth. The strong longitudinal electric field of the plasma wake accelerates this ``witness'' beam.  As the performance of a PWFA depends strongly on the plasma density profile~\cite{Xu2016, Floettmann2014, Ariniello2019, Ariniello2022}, control and understanding of the plasma source are essential to interpret the outcome of experiments and optimize the properties of the plasma source.

A typical PWFA plasma source is a long, narrow filament, 10's to 100's of micrometers in diameter and 10's of centimeters in length, with a core density of $10^{16-18} \ {\rm cm^{-3}}$. One technique to form a suitable plasma is to laser-ionize a gas using an optic with a long depth of focus, such as an axicon lens or diffractive optic~\cite{Honkanen1998, Oua, RobertThesis, lee2019}. In such a plasma source, an $\mathcal{O}(10 \,{\rm TW})$, ultrashort laser pulse is focused into a volume of gas, such as Hydrogen, Helium, or Lithium, ionizing a thin filament prior to the arrival of the electron beams. Characterization of the plasma source is often performed in the absence of the electron beams. One common technique is to look at the light emitted by the plasma.

Multiple mechanisms with varying time scales contribute to the plasma light emission process of these thin laser-ionized plasma sources. The plasma is locally formed on the time scale of the ionizing laser pulse duration, which is around $100 \ {\rm fs}$. The plasma electrons thermalize within $10 \ {\rm ps}$ of formation. Then, the plasma expands quickly outward within a few nanoseconds while the neutral gas diffuses inward, followed by a slower diffusion and thermalization phase between the plasma and the neutral gas that lasts 10's to 100's of nanoseconds. During this time, plasma electrons collide with neutral atoms, exciting them, while also colliding with plasma ions and recombining. Both of these mechanisms contribute to plasma light emission. The latter will also lead to the eventual neutralization of the plasma over a few hundred microseconds time scale~\cite{Kewtont}. 

The dynamics of plasma expansion have been studied through simulations and experiments~\cite{Shalloo2018, Milos2016, Capitelli2004}, and the electron thermalization time scale has been demonstrated in Ref.~\onlinecite{Zhang2020, Zhang20201}. Experimental and theoretical studies of plasma neutralization through recombination are presented in Ref.~\onlinecite{Baravian1972, Bates1962}. However, there remains a gap in understanding the light emission process in thin, laser-ionized plasmas, such as PWFA plasma filaments. The amount of light emitted depends on the electron temperature and the number densities of the plasma and neutral gas. Determining the relative dominance of excitation versus recombination is a complex question influenced by the plasma's geometry. In this work, we show an analytical and computational model of the plasma light emission process of a thin laser-ionized Helium plasma and confirmed by the experiment. 

Our measurements of plasma light emission are taken using an inexpensive camera. These cameras are often used to verify plasma formation in PWFA experiments. For example, many of the experiments at the Facility for Advanced Accelerator Experimental Tests-II (FACET-II) at SLAC National Accelerator Laboratory rely on cameras to observe the laser-ionized plasma source~\cite{Joshi2018, Deng2019, Adli2016, Miguel2022, Amorim2019, Corde2016, Doss2019}. Cost-effective cameras, such as Gigabit Ethernet (GigE) machine vision cameras, are typically used in PWFA experiments because the high radiation environment in the accelerator housing leads to rapid cycles of camera failure and replacement. Unfortunately, these cameras have a large trigger timing jitter ($\sim 10 \ {\rm \mu s}$) and long exposure time ($\sim 10 \ {\rm \mu s}$) compared to the short time scales (10's-100's ${\rm ns}$) of plasma light emission, making images from these kinds of cameras integrate over various dynamics described previously. In order to verify our model using these cameras, we demonstrate a novel technique to measure the time-resolved light signal using these low-cost cameras.


Our three step light emission model: plasma formation, plasma expansion, and plasma light emission is presented in section~\ref{sec:modeling}. In section~\ref{sec:exp}, we introduce our experimental setup where a laser-ionized Helium plasma is viewed by a GigE machine vision camera. In section~\ref{sec:meth}, we present a novel technique for studying plasma light emission with 1 ns resolution using a low-cost GigE camera. In section~\ref{sec:res}, we demonstrate that the experimental data confirms our Helium plasma light emission model. Furthermore, we show the broad applicability of this diagnostic tool in PWFA experiments by demonstrating that a simple theoretical model can describe the intensity of the plasma light in a time-integrated image, permitting estimation of the initial plasma parameters with a single image of the time-integrated plasma light.

\section{\label{sec:modeling}Modeling}

This section outlines a workflow for modeling the laser-ionized plasma formation, expansion, and light emission. We demonstrate the entire workflow using a single set of parameters (laser pulse duration: $\tau$= 50 fs, focusing optic: $1^{\circ}$ axicon lens, laser energy: $E$= 160.33 mJ, gas pressure: $P$= 118 mbar) and compare the time-resolved simulated plasma light emission pattern with experimental data. We also conducted multiple plasma formation simulations for a range of parameters ($E$= 140-175 mJ and $P$= 20-80 mbar). The simulated electron temperatures were used to calculate plasma light emission, which is compared to experimental measurements in Section~\ref{sec:res}.


\subsection{\label{sec:plasmaform}Plasma Formation}

\begin{figure}[t]
\includegraphics[width=1\columnwidth]{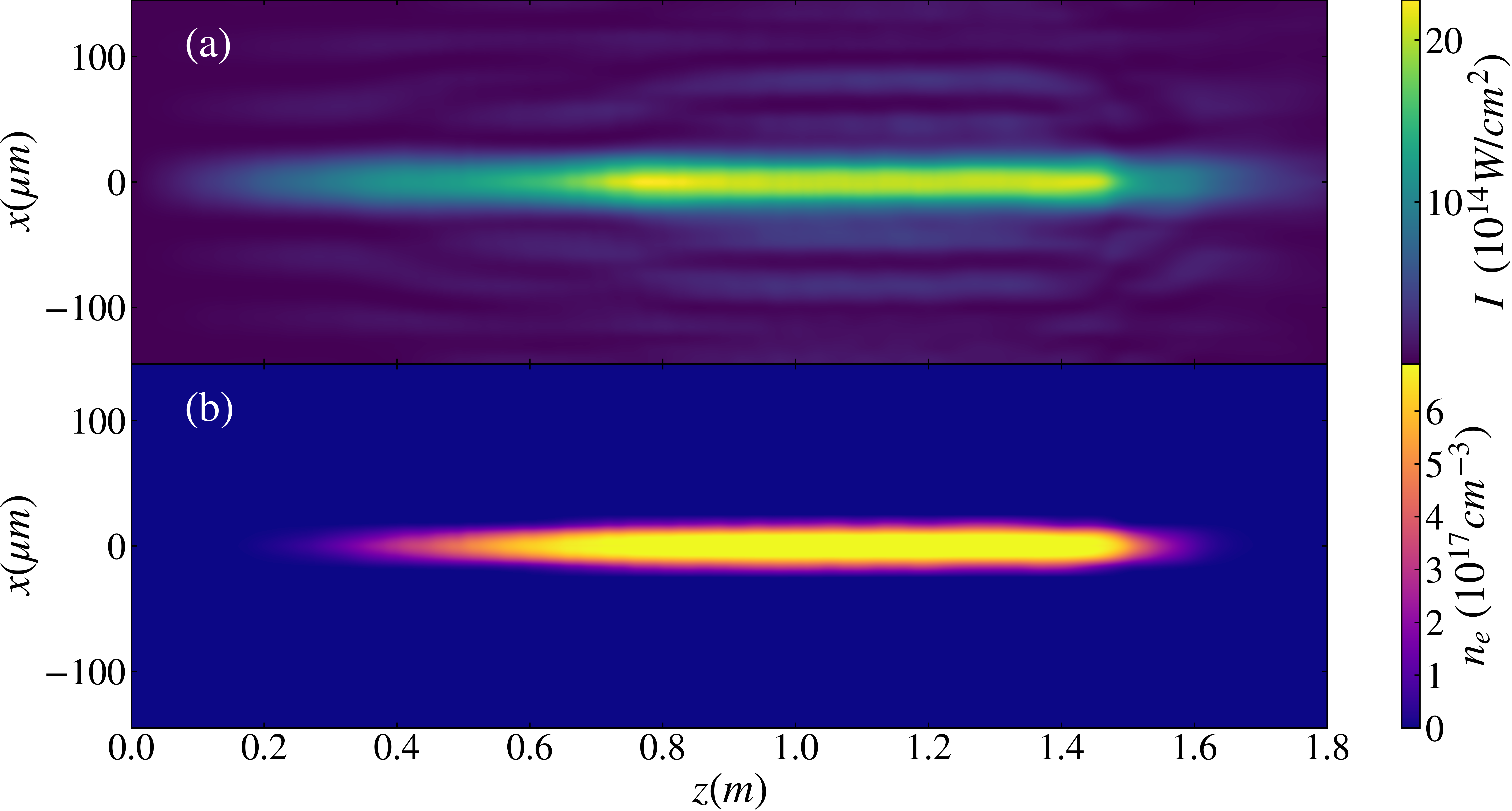}
\caption{\label{fig:laserplasma} Simulated formation of an axicon plasma using a split-step Fourier code, with an initial laser pulse energy of 175~mJ in a Helium gas pressure of 28~mbar, corresponding to a fully ionized plasma of $n_e= \ 6.9\times 10^{17} \ {\rm cm^{-3}}$. The axicon lens ($\alpha= 1 ^{\circ}$) is positioned at $z=0$. (a) illustrates the laser intensity at the middle of the pulse on the $y=0$ plane. (b) shows the plasma density on the $y=0$ plane. Because of the long Bessel focus from the axicon lens, the full ionization regime extends to up to 1 meter.}
\end{figure}

In a laser-ionized PWFA plasma source, ionization occurs through field ionization, typically in the tunneling regime where the electric field from the laser distorts the atomic potential allowing an electron to tunnel out of the atomic barrier. This process is well described by the ADK model \cite{keldysh1965ionization}. Particle-in-cell (PIC) simulations are able to capture this process, however, simulating the ionization process over a 5 ns window (axicon focus of 1.5 m) while resolving the laser period would require an impractical amount of computational resources. As an alternative, we use an in-house code to simulate the 3D plasma profile produced by the laser. The ADK model is used to calculate the ionization rate; while a the split-step Fourier (SSF) algorithm~\cite{stoffa1990split} takes into account refraction of the rear of the laser pulse due to the presence of plasma ionized by the front of the laser pulse. Figure~\ref{fig:laserplasma} shows an example of the simulated laser intensity and plasma density profiles formed by an axicon lens by our SSF code. This code, however, cannot provide information about the initial plasma electron temperature. To accommodate this, we use a PIC code, VSim~\cite{nieter2004} provided by Tech-X, to simulate ionization of a typical localized region along the laser ionization path, importing the local laser pulse profile retrieved from the SSF code. We then extract the plasma electron temperature from the resulting kinetic energy distribution of the electrons in the PIC simulation.

\begin{figure}[]
\includegraphics[width=1\columnwidth]{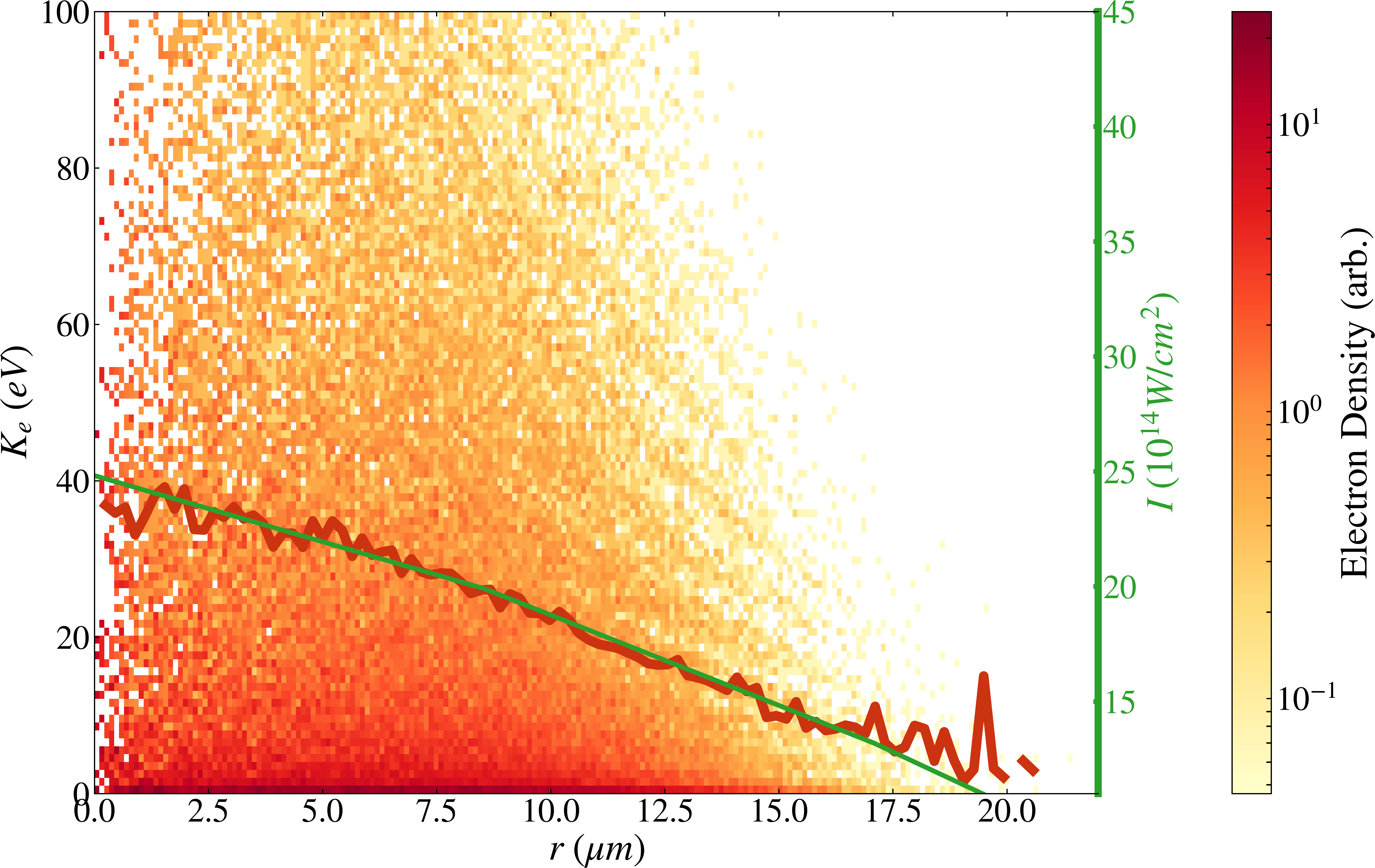}
\caption{\label{fig:PIC} The plot illustrates the plasma electron energy histogram as a function of the radius, ionized by a $175 \ {\rm mJ}$ pulse in Helium gas with a density of $6.9\times 10^{17} \ {\rm cm^{-3}}$, simulated by a PIC simulation. The azimuthally averaged radial electron energy is represented by the red curve and the ionizing laser intensity is represented by the green curve. The azimuthally averaged radial electron energy follows the intensity of the ionization laser.}
\end{figure}

For all PIC simulations in this work, 1 particle per cell is used with a time step of $dt= 3.25\times 10^{-17} \ {\rm s}$ and a spatial increment of $dx= 7\times 10^{-8} \ {\rm m}$. The transverse size of the simulation box is approximately 14 times the full width at half maximum (FWHM) of the laser's spatial profile, and the temporal dimension is roughly 4 times the FWHM duration of the laser pulse. Modeling ionization through a combination of the SSF and PIC simulations allows us to acquire the plasma density and plasma electron temperature with reasonable computational resources. Figure~\ref{fig:PIC} shows a histogram of the electron kinetic energy distribution as a function of radius. It also demonstrates that the intensity of the ionization laser profoundly influences the azimuthally averaged electron energy. As we will show, our experimental observations confirm that the method described above is a reliable and relatively inexpensive way to model a laser-ionized PWFA plasma source.


\subsection{\label{sec:plasmaexpa}Plasma Expansion}

\begin{figure}
\includegraphics[width=1\columnwidth]{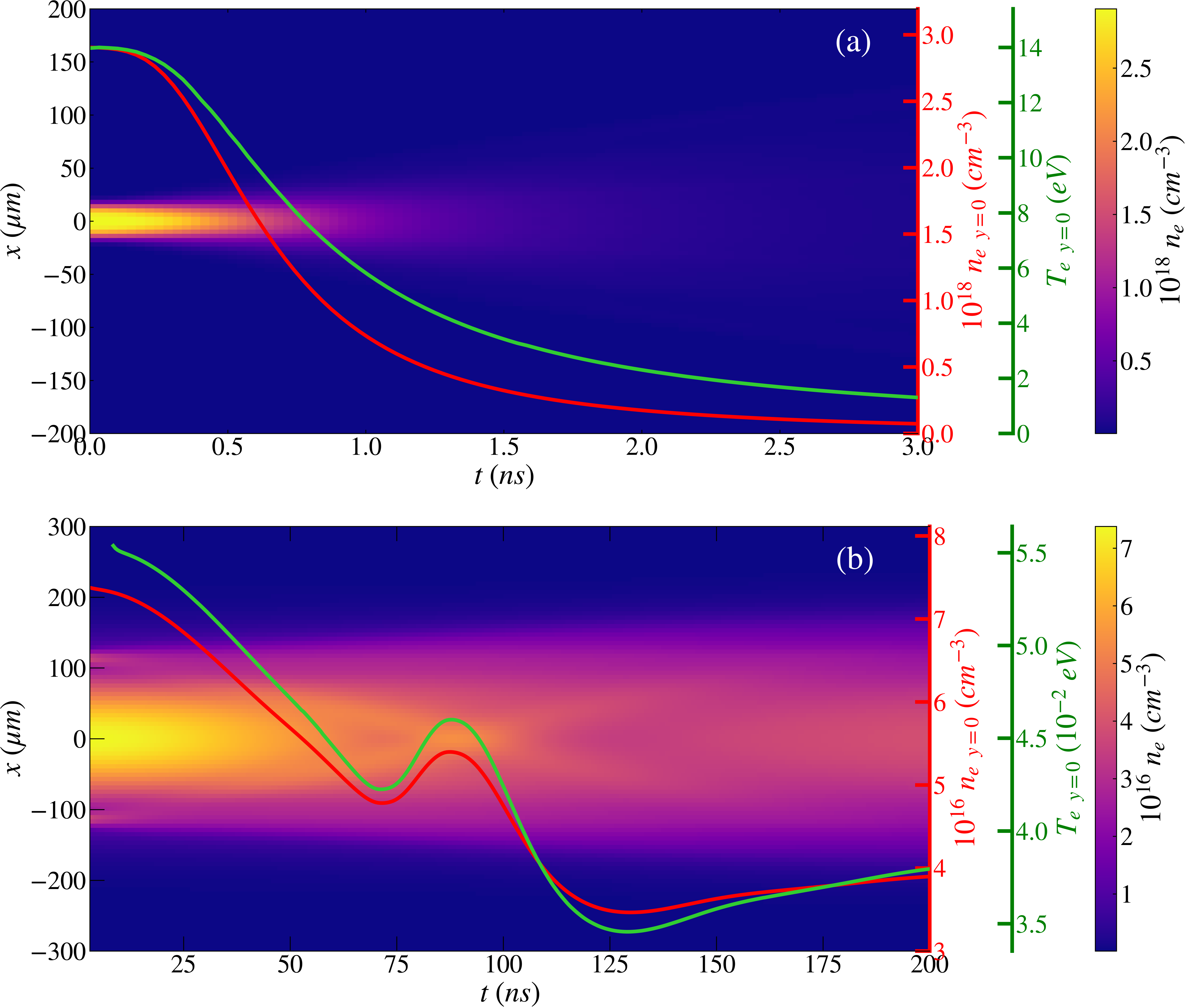}
\caption{\label{fig:expand} Simulated plasma expansion using fluid codes. (a) shows the plasma expansion on the $y=0$ plane within the first three nanoseconds simulated by a single-fluid MHD model. (b) shows the plasma expansion on the $y=0$ plane from the 3~ns to 200~ns, including plasma-neutral collision effects. In both figures, the red and green curves show the plasma density ($n_e$) and plasma temperature ($T_e$) at $x=0$, respectively. Note that the first 5~ns (numerical thermalization time) of the green curve ($T_e$) in (b) is excluded from the plot.}
\end{figure}

The plasma electrons thermalize on the order of the Spitzer electron self-collision time, given by $\tau \approx (1.40/(8 \pi r_e^2 c^4 n_e \ln \Lambda))(3k_B T_e /m_e )^ {3/2}$ where $r_e$ is the classical electron radius, $c$ is the speed of light, $n_e$ is the plasma electron density, $\ln\Lambda$ is the Coulomb logarithm, $k_B$ is the Boltzmann constant, $T_e$ is the average electron temperature, and $m_e$ is the mass of electron~\cite{spitzer1962physics}. For example, for a plasma with $k_BT_e= 10 \,{\rm eV}$ and $n_e= 1\times 10^{17}\,{\rm cm^{-3}}$, $\tau \approx 9 \,{\rm ps}$. This fast thermalization process has been measured in Ref.~\onlinecite{Zhang2020, Zhang20201}. The thermalized electron temperature of the plasma is $k_B T_e= 2/3<E_k>$, where $<E_k>$ is the average electron kinetic energy immediately after ionization. The expansion of a laser-ionized plasma source can be modeled using fluid simulations, as demonstrated in Ref.~\onlinecite{Shalloo2018}. Using the plasma density profile and the plasma electron temperature acquired from previous steps, we model the plasma expansion using Tech-X’s fluid simulation software, USim. USim is an Eulerian computational fluid dynamics code optimized for plasma fluid simulations by solving the Magnetohydrodynamic (MHD) equations~\cite{Brio1988}. We simulate the first $3 \,{\rm ns}$ of the expansion using a two-temperature, single-fluid MHD diffusion code, where the ion temperature is $0.025\,{\rm eV}$ ($300\,{\rm K}$) and the average electron temperature is taken from the PIC simulation, $<T_e>\sim 13\,{\rm eV}$. The simulated expansion is shown in figure~\ref{fig:expand}(a). For the subsequent expansion from $3 \,{\rm ns}$ until $t=200 \,{\rm ns}$, plasma-neutral collisions are accounted for using a two-fluid model that calculates the mass diffusion, energy exchange, and temperature exchange between two species (plasma and neutral)~\cite{Meier2012, Chaplin2015}. The subsequent expansion is shown in Fig.~\ref{fig:expand}(b). It has been tested that the results of the second simulation period are relatively insensitive to the exact choice of the transition time, so long as the change in density is relatively small. As shown by the red curve (density lineout) in Fig.~\ref{fig:expand}(a), the expansion (decay) reaches $1/e^2$ before $3 \,{\rm ns}$. Simulating the expansion in two steps simplifies the complexity of the model and allows for a higher temporal resolution during the initial rapid expansion.

\subsection{\label{sec:plasmalight}Plasma Light Emission}

Light emission from plasma can be generated by two primary processes: electron-neutral collisional excitation and plasma electron-ion recombination. Both processes are intricate and require Monte Carlo atomic electron tracking models~\cite{vahedi1995monte} to simulate exactly, which is both theoretically and computationally demanding. To simplify the problem, we focus on the most probable transition process that emits visible/NIR light in either process, since that is what Complementary Metal-oxide Semiconductor (CMOS) GigE camera can detect.

In the collisional excitation model, the electrons have a significantly higher temperature ($T_e \sim 15 \,{\rm eV}$) compared to the ions and neutrals ($T_i \approx T_n \sim \,{\rm 0.025 eV}$). Therefore, electron-neutral collisions dominate while ion-neutral and neutral-neutral collisions are negligible. We also assume that all neutral atoms are in their ground state before the collision. In a Helium atom, $3d-2p$ de-excitation has the strongest persistent lines. The dipole-allowed transitions from the ground state occur from the $1s$ ($l=0$) state to the $p$ ($l=1$) states. Of these, excitation to the $4p$ state has the largest cross-section among allowed excited states that can generate a visible/NIR photon from a subsequent $3d-2p$ de-excitation. Therefore, we choose this $1s-4p-3d-2p$ excitation/de-excitation process to calculate the lower bound of the visible photon emission rate via a single excitation path. Appendix A presents a more detailed explanation of this model. To calculate the local excitation rate per unit volume, we start with the collision rate for one electron with a population of neutral particles\cite{Theodosiou1987, Boffard2004, Ralchenko2008},

\begin{eqnarray}
\upsilon_{en}= n_n \sigma(v)_{1s-4p} \ v,
\label{eq:SingleColR}
\end{eqnarray}
where $n_n$ is the local neutral density, $\sigma(v)_{1s-4p}$ is the electron-impact excitation cross-section from the 1s state to the 4p state, and $v$ is the relative velocity between the electron and the neutrals, which is approximated as the electron velocity ($v_e>> v_n$). The probability density function for a given speed ($v=|\vec{v}|$) for a Maxwellian population of electrons is
\begin{eqnarray}
f(v)= \left(\frac{m_e}{2 \pi k_B T_e}\right)^{3/2} \ 4\pi v^2 \exp\left({-\frac{m_e v^2}{2 k_B T_e}}\right),
\label{eq:Maxwellian}
\end{eqnarray}
where $\int_{0}^{\infty} f(v) dv = 1$. The average collision rate per electron is
\begin{eqnarray}
&&<\upsilon_{en}>= \frac{\int_{0}^{\infty} f(v) n_n \sigma(v) v dv}{\int_{0}^{\infty}f(v) dv} \nonumber\\
&&= n_n 4 \pi \left(\frac{m_e}{2 \pi k_B T_e}\right)^{3/2} \int_{0}^{\infty} \exp\left({-\frac{m_e v^2}{2 k_B T_e}}\right) \sigma(v) v^3 dv.
\label{eq:NorMaxwellian}
\end{eqnarray}
This average collision rate can be written in terms of the kinetic energy, $K$:
\begin{eqnarray}
<\upsilon_{en}> = &&\frac{8 \pi n_n}{m_e^2} \left(\frac{m_e}
{2 \pi k_B T_e}\right)^{3/2}
\nonumber \\
&&\int_{0}^{\infty}
\exp\left({-\frac{K}
{k_B T_e}}\right)
\sigma(K) K \ dK.
\label{eq:NKiMaxwellian}
\end{eqnarray}
Thus, the excitation rate per unit volume at location $\vec{r}$ and time $t$, for a population of electrons colliding with a population of neutrals is $n_e<\nu_{en}>$, which depends on $\vec{r}$ ant $t$ through $n_e$, $n_n$, and $T_e$.
%

After the collisional excitation, some excited electrons in the $4p$ state spontaneously de-excite to the $3d$ state, and then to the $2p$ state, yielding a detectable photon emission. The decay transition probability from state $i$ to state $j$ is
\begin{equation}
P_{ij}= \frac{A_{ij}}{\sum_{k} A_{ik}},
\label{eq:TransProb}
\end{equation}
where $k$ indicates all dipole-allowed final transition states from initial state $i$, and $A$ is the transition rate, here taken from Ref.~\onlinecite{Theodosiou1987}.

The transition probability from the $4p$ state to the $3d$ state is $P_{4p-3d}\approx 1.2\times 10^{-3}$, and from the $3d$ to the $2p$ state is $P_{3d-2p} \approx 1$. The visible photon emission rate is thus
\begin{eqnarray}
\frac{dn_{\gamma exc}(\vec{r}, t)}{dt}= n_e(\vec{r}, t)<\upsilon_{en}(\vec{r, t})> \nonumber P_{4p-3d} \ P_{3d-2p}.
\label{eq:photonExc}
\end{eqnarray}

We then proceed to work out an analytic prediction of the peak photon emission $(x=0, y=0)$ scaling as a function of the initial plasma density and temperature with a few assumptions explained in Appendix B:

%
\begin{eqnarray}
\Gamma= &&\int_{0}^{\infty} n_e(x=0, y=0)<\upsilon_{en}(x=0, y=0)> dt \nonumber\\
= &&C \ n_0^2 \sqrt{R k_B T_0} \ \exp\left({-\frac{K_{th}}{Rk_BT_0}}\right),
\label{eq:ana}
\end{eqnarray}
where $C$ is scaling constant left as a free parameter to fit to the data, $n_{0}$ and $T_{0}$ are the initial electron density and temperature at $x=0, y=0$, respectively, $K_{th}$ is the free electron kinetic energy threshold for exciting a ground-state atom to the $4p$ state, and $R = 0.09$ is the empirically acquired temperature decay ratio; that is, the ratio between the initial temperature and the final temperature after the initial fast expansion as shown in Fig.~\ref{fig:expand} (a). This value was obtained from our simulation and may vary in other circumstances, though it agreed well with the experimental data over a relatively wide range of initial plasma parameters. The detailed derivation of \ref{eq:ana} can be found in Appendix B. 

The other possible light emission process, recombination, can occur through three different primary modes: three-body recombination, radiative recombination, and dielectronic recombination~\cite{Hahn1997}. Radiative recombination dominates the production of visible photons~\cite{Drawin1975}, wherein a single free electron recombines with a singly ionized ion into one of its high Rydberg states, yielding a highly-exited Helium atom. The rate of electron-ion radiative recombination can be approximated as follows~\cite{NRL}:
%
\begin{equation}
\alpha_r= 2.7 \times 10^{-13}
Z^2 \ T_e[eV]^{-1/2}
\ \ [{\rm cm^3/s}],
\label{eq:NRL}
\end{equation}
where $Z=1$ is the charge state of He$^+$.
The recombination rate~\cite{chen1984introduction} is, therefore, 
\begin{equation}
\frac{dn_{\gamma {\rm rec}}}{dt} = \alpha_r n_e^2,
\label{eq:ChenBook}
\end{equation}
assuming quasi-neutrality.

Because it is more energetically favorable, we assume that most of the recombined electrons start in a highly-excited state and then eventually de-excite to the ground state in a radiative process. To estimate an upper bound of the detectable photon emission rate from a single recombination event, we look at the strongest visible/NIR line in Helium atoms, the $3d-2p$ transition. We then assume that electrons in high Rydberg states eventually go through the dominant transition process, $4f-3d-2p$ ~\cite{Theodosiou1987, Ma2011}. The details of the model are discussed in Appendix A. The de-excitation transition probability for $4f$ states to $3d$ states is $P_{4f-3d} \approx 1$. The local visible photon emission rate from electron-ion recombination is thus
\begin{equation}
\frac{dn_{\gamma rec}(\vec{r}, t)}{dt}=
\alpha_r n_e(\vec{r}, t)^2 
\ P_{4f-3d} P_{3d-2p}.
\label{eq:photonRecombo}
\end{equation}
Note that \ref{eq:photonRecombo} is predicated on the assumption that all recombination events lead to the emission of a detectable photon, which is not strictly true, but does provide an upper-bound estimation of the recombination photon emission rate.



The collisional excitation and recombination rates per unit volume are orders of magnitude less than the electron density, so their effects on the expansion of the plasma are negligible. This allows us to model the time-resolved plasma expansion first and then numerically estimate the visible photon emission density from excitation and recombination using Eq.~\ref{eq:photonExc} and Eq.~\ref{eq:photonRecombo} at each moment in time, $t$, and at every position, $\vec{r}$. The cross-section in Eq.~\ref{eq:SingleColR} is numerically interpolated from Ref.~\onlinecite{stone2002}. Our simulation indicates that the visible/NIR photon emission rate from the collisional excitation is three orders of magnitude greater than that from recombination. Recall that in our estimation, the photon emission rate arising from collisional excitation is a lower bound, while that from plasma recombination is an upper bound. Thus, even with the simplifications introduced in our model, the assertion that collisional excitation principally dominates the light emission process remains valid. Figures~\ref{fig:4plots} (a) illustrates the time-resolved simulated photon emission pattern from an expanding Helium plasma. 

\section{\label{sec:exp}Experimental Setup}


\begin{figure*}
\centering
\includegraphics[width=0.8\textwidth]{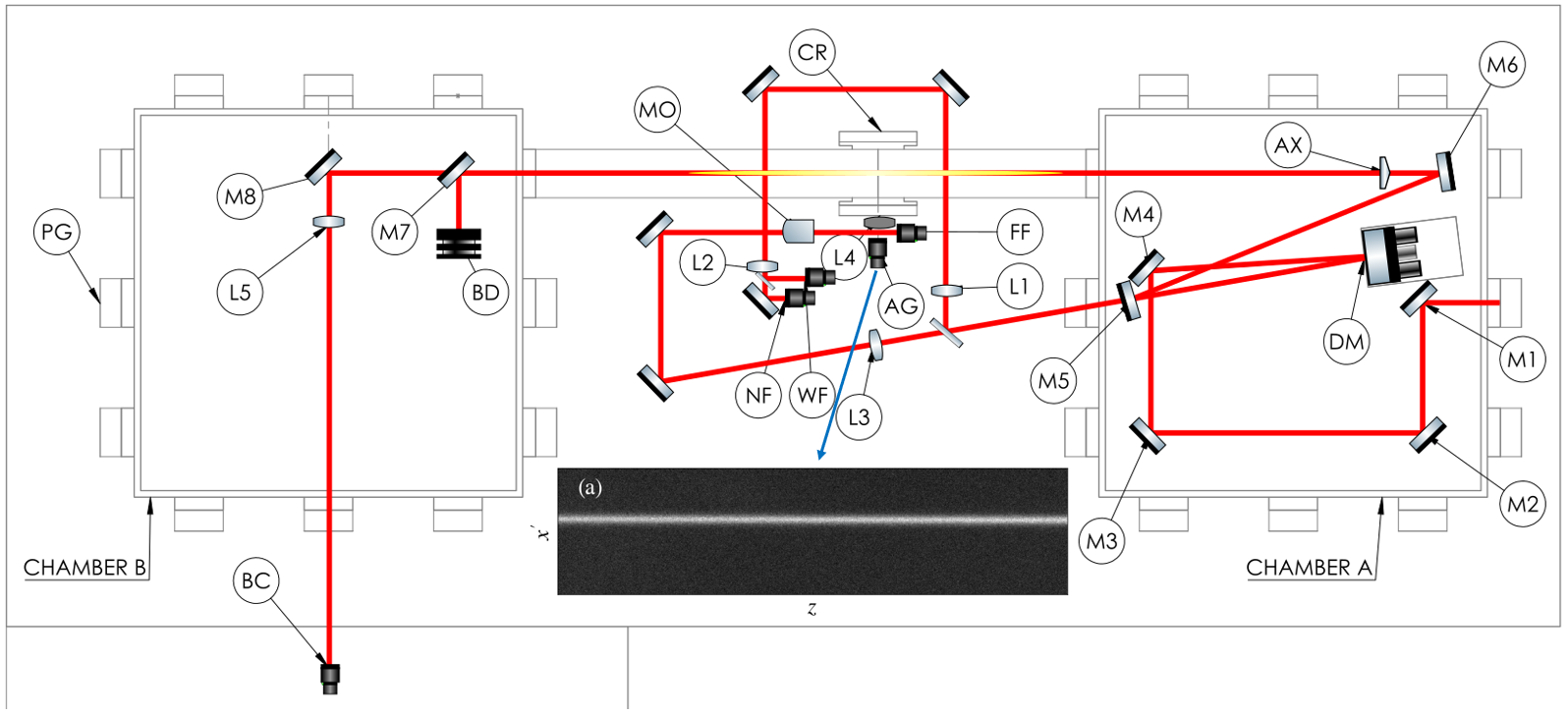}
\caption{\label{fig:setup} Experimental setup. 50~fs laser pulses generated by a Ti:Sapphire laser system enter Chamber A through the rightmost window in this figure, bouncing off mirrors M1, M2, M3, M4, and the deformable mirror (DM). The leakage light from M5 exits the vacuum chamber and splits into three paths to the near-field camera (NF), far-field camera (FF), and the Shack-Hartmann wavefront sensor (WF) to monitor the laser position, pointing, and wavefront. (L1, L2: lenses; MO: microscope objective.) The main laser beam reflects off mirrors M5 and M6, passing through a $1 ^{\circ}$ axicon lens (AX), which produces a Bessel focus within the vacuum pipe connecting Chamber A and Chamber B. The Bessel focus is centered at the six-way cross (CR), as denoted by the yellow region indicating plasma formation. The laser beam is terminated at the beam dump (BD), while the leaked light from mirror M7 is used for imaging the Bessel focus by a single lens (L5) to a camera (BC). On the diagnostic side, a commercial macro lens lens (L4) is used to image the plasma light onto a camera (AG). The vacuum pressure is measured shot-to-shot using a pressure gauge (PG) on the figure's far left. (a) shows an example of the raw plasma light image taken by camera AG.}
\end{figure*}

The plasma channel is ionized by a $50\,{\rm fs}$, $450\,{\rm mJ}$ pulse from a terawatt-class Ti:Sapphire laser system. The wavefront of the pulse is optimized with a deformable mirror before being sent through an axicon lens ($\alpha= 1 ^{\circ}$) that generates a non-diffracting Bessel-beam focus over a length of $\sim150\,{\rm cm}$. The experimental vacuum chamber is filled with Helium gas, creating the plasma source within. The vacuum system consists of two vacuum chambers connected with vacuum pipes and a six-way vacuum cross as shown in Fig. 4. The plasma is formed inside the vacuum pipe and viewed through an optical window. The gas pressure is monitored by a vacuum gauge and recorded on a shot-by-shot basis. The laser intensity profile is recorded upstream and downstream of the axicon lens. The former is used as an input parameter for the simulation, and the latter is used to monitor the Bessel focus profile.

The diagnostic system views the plasma light at 1 m downstream of the axicon. A commercial camera lens (AT-X M100 PRO D Macro from Tokina, $f=100 \ {\rm mm}$) is installed immediately outside the vacuum window. A $6$-OD notch filter (central wavelength= $785 \ {\rm nm}$, FWHM= $33 \ {\rm nm}$) filters out most of the scattered laser light that appears as a background to the plasma light. The imaged plasma light is recorded with a CMOS GigE camera chip (Sony IMX265), forming the primary source of data collected in this experiment. 


The camera is externally triggered by a trigger pulse train from a Signal Delay Generator (SDG), which is synchronized with the Ti:Sapphire laser system. The timing of the laser system is precise to the sub-picosecond scale; the SDG is precise to the 10's of picoseconds scale. However, the CMOS GigE camera has a significant internal timing jitter between the arrival of the trigger signal to the initiation of the exposure time, on the scale of $10 \ {\rm \mu s}$. The shortest integration time for this type of camera is on the order of $10 \ {\rm \mu s}$, whereas the dynamic timescale of light emission from the plasma is 10’s to 100's of nanoseconds. Therefore, this type of camera always captures a time-integrated signal of the plasma light that is much longer than the dynamic timescale.


Despite the limitations of CMOS GigE cameras, their cost-effectiveness makes them a preferred choice in high-radiation environments. For example, in the FACET-II accelerator tunnel at SLAC National Accelerator Laboratory, approximately 10 GigE cameras are employed solely to monitor the laser and plasma source. During operations, FACET has reported the loss of more than 20 cameras per year due to high radiation exposure. Therefore, cost-efficient CMOS cameras (tyipcally a few hundred dollars each) are far more suitable than would be a fast (nanosecond) gated camera, which can cost tens of thousands of dollars. In Section~\ref{sec:meth}, we introduce an innovative analysis method to extract time-resolved plasma light emission data from time-integrated measurements using inexpensive GigE cameras.

\section{\label{sec:meth}Analysis Methodology}

\subsection{\label{sec:stat}Statistical Analysis}
As explained in the previous section, CMOS GigE cameras are preferable in high-radiation environments but are conventionally unsuitable for high-timing-precision measurements. In this section, we discuss an inventive statistical analysis to extract time-resolved light emission from multi-shot time-integrated data.

\begin{figure}[h]
\includegraphics[width=1\columnwidth]{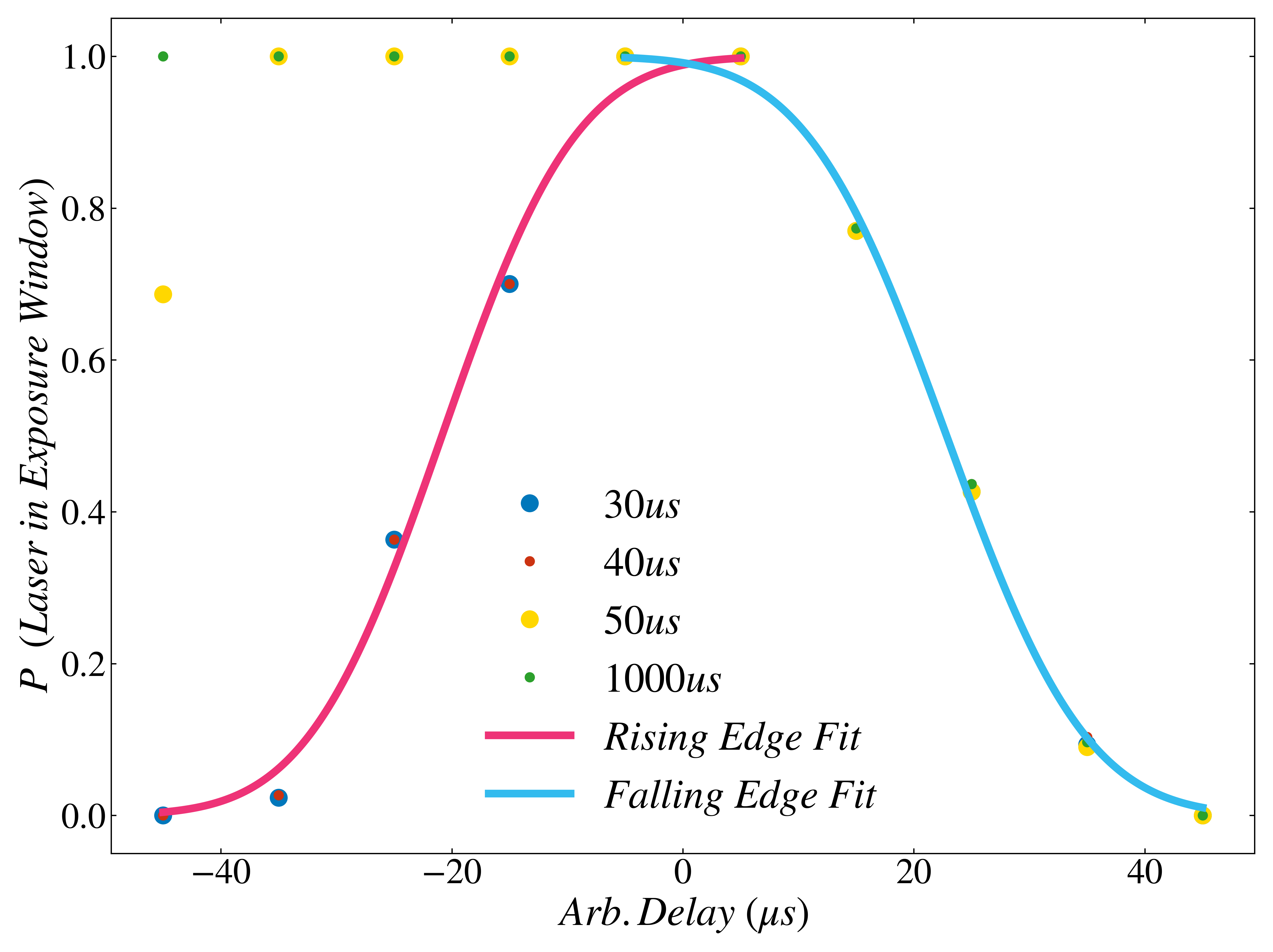}
\caption{\label{fig:PLinW} Measurement of the probability of a laser pulse appearing within the camera’s exposure time at various relative delays, measured with 30, 40, 50, and $1000\,{\rm \mu s}$ exposure settings. Jitter of the camera's exposure start time and end time cause the falling and rising ramps to be continuous functions instead of step functions. Repeatable measurements of the falling edge at the various exposure settings (30, 40, 50, and $1000\,{\rm \mu s}$) confirm that altering the exposure duration does not affect the start-time jitter. The magenta and cyan curves show Eq.~\ref{eq:cumulative} fitted to the rising and falling ramps, respectively.}
\end{figure}

\begin{figure}
\includegraphics[width=1\columnwidth]{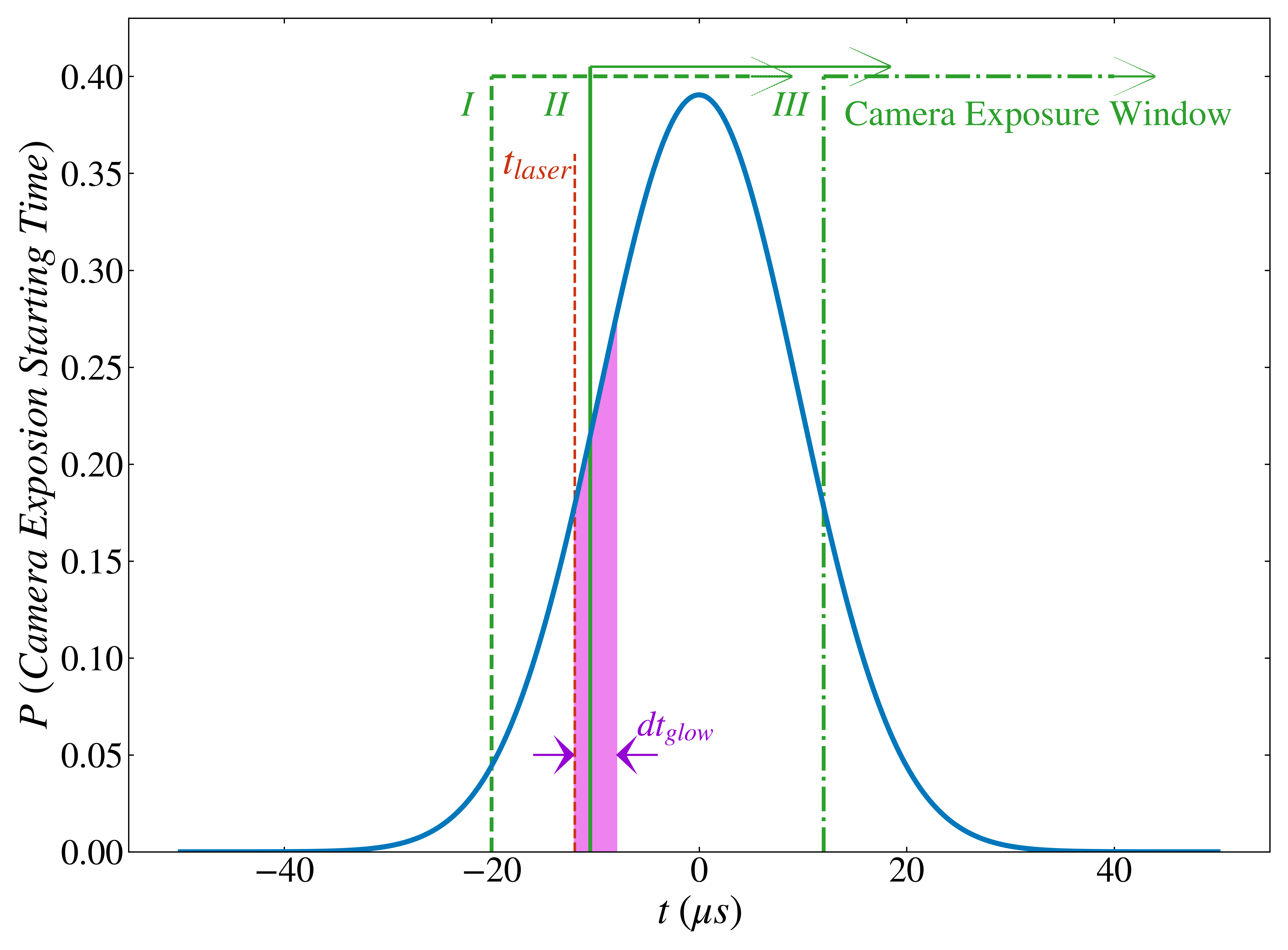}
\caption{\label{fig:luckshot} A conceptual diagram illustrating the relationship between camera jitter, laser arrival time, and plasma light time scale. The blue curve represents the camera starting time jitter distribution measured and calculated by Eq.~\ref{eq:cumulativeRel}. The red dashed line indicates the laser arrival time. Following the laser arrival, if a plasma is formed, the plasma light continues to glow for an extended duration, $dt_{glow}$, until the intensity is undetectable. This is represented as the purple area shown in the figure, though the scale of $dt_{glow}$ is greatly exaggerated for illustration. When the camera exposure starts at time $I$ (dashed green line), the captured image contains both laser and plasma light. When it starts at time $III$ (dot-dashed green line), no light is captured in the image. However, when it starts at a propitious timing, {\it e.g.}, at time $II$ (solid green line), the camera exposure begins after the arrival of the laser but before the plasma light dissipates, resulting in the capture of only the plasma light without the laser light.}
\end{figure}

The first step is quantifying the camera-trigger-jitter distribution, which is device-dependent. Physically, a trigger pulse is sent at time $t_{\rm SDG}$ to the GigE camera. After processing in the camera chip, the exposure time begins at $t_{\rm start}$. This delay between $t_{\rm SDG}$ to $t_{\rm start}$ varies shot-to-shot; we call this the camera-trigger-jitter, which follows a distribution $J(t)$. The precision of the laser arrival time at the target plasma location ($\mathcal{O}(10 \,{\rm ps})$) is orders of magnitude smaller than the camera-trigger-jitter ($\mathcal{O}(10 \,{\rm \mu s})$). As a result, the laser arrival time can serve as a fiducial signal to investigate $J(t)$. The diagnostic camera is first synchronized to the laser arrival-time, so that the laser pulse always arrives within the duration of the camera exposure time and the laser scattered light is captured by the camera in every shot. This starting time is defined as $t= 0$ in Fig.~\ref{fig:PLinW}. The trigger signal is then advanced/delayed away from $t= 0$. Because the exposure starting-time (and ending-time) jitters, the laser pulse will sometimes fall outside the exposure window as the delay time is scanned, resulting in images without laser signal. Eventually, the trigger advance/delay is great enough that none of the images taken capture the laser. The observed probability of a laser pulse falling inside the camera exposure window as a function of advance/delay time, $j(t)$, is plotted in Fig.~\ref{fig:PLinW}. Note that the falling edge in this figure corresponds to a relative {\em delay} of the SDG trigger, meaning that the laser arrives {\em ahead} of the camera exposure’s starting-time. This implies that the falling edge of $j(t)$ is the cumulative distribution function of the camera-trigger-jitter distribution, $J(t)$, 

\begin{equation}
j(t)= \int_{-\infty}^{t} J(t)dt.
\label{eq:cumulativeRel}
\end{equation}
The cumulative distribution function of a normal distribution is
\begin{equation}
j(t)= \frac{1}{2}\left( 1+ \text{erf} \left( \frac{t- \mu_J}{\sigma_J \sqrt{2}}\right)\right),
\label{eq:cumulative}
\end{equation}
where $\mu_J$ is the mean of $J(t)$, $\sigma_J$ is the deviation of of $J(t)$, and erf is the error function.
The rising and falling edge data presented in Fig.~\ref{fig:PLinW} is fitted with Eq.~\ref{eq:cumulative} and the camera-trigger-jitter distribution, $J(t)$, is thus determined from Eq.~\ref{eq:cumulativeRel}, assuming 
normal distribution and $\sigma_J$ is given by the fitting, and plotted in Fig.~\ref{fig:luckshot}


The observable plasma light lasts for an extended duration of time $dt_{\rm glow}$. Most of the images captured by the camera either include both the plasma light and background light from the laser pulse, or neither. (Note: Though the notch filter removes much of the scattered laser light, some always remain visible on the camera.) For certain trigger delay times, however, it is possible for the camera to capture a portion of the plasma light while the prompt laser light falls outside of the camera’s exposure window. This happens when the exposure begins after the laser arrives but before the plasma light ends. The longer the plasma light decay time, $dt_{\rm glow}$, the more these ``lucky shots'' accumulate.

The purple area under the curve in Fig.~\ref{fig:luckshot} represents these ``lucky shots'' (shots with plasma light but without the laser signal). The ratio of ``lucky shots'' to all recorded shots $R$ is expressed as
%

\begin{eqnarray}
R\ = 
\int_{t_{\rm laser}}
^{t_{\rm laser}+dt_{\rm glow}}
J(t- \ t_{\rm SDG}) \,dt 
\end{eqnarray}
where $J(t-\ t_{\rm SDG})$ is the camera-trigger-jitter distribution centered about a given SDG trigger time $t_{\rm SDG}$, and $t_{laser} \equiv t_{SDG_0}$, where $t_{SDG_0}$ is the timing $t=0$ at Fig.~\ref{fig:PLinW}. 
Accordingly, we solve for $dt_{\rm glow}$, giving 
\begin{eqnarray}
dt_{glow}|_{t_{SDG}}= 
\left. \frac{-J(t_{\rm laser})
+\sqrt{J(t_{\rm laser})^2-2\ I\ J'(t_{\rm laser})}}{J'(t_{\rm laser})}\right| _{t_{SDG}}, 
\label{eq:dt}
\end{eqnarray}
where $J'$ denotes the time derivative of $J$, and $I= R{\int_{-\infty}^{\infty} J(t- \ t_{\rm SDG}) \,dt}$. Using Eq.~\ref{eq:dt}, we can measure the characteristic plasma emission time scale without a nanosecond gated camera, so long as we record a statistically sufficient number of images with our GigE camera. 

\subsection{\label{sec:DataPro}Data Processing}

In this subsection, we explain how the raw experimental data is processed before performing the statistical analysis described in the previous subsection. We collected data at various SDG delays across a $120 \,{\rm \mu s}$ interval with $10 \,{\rm \mu s}$ increments. For each SDG delay, we set the camera exposure time to $30$, $40$, $50$, and $1000 \,{\rm \mu s}$. We captured $300$ shots for each combination of SDG delay and exposure time, resulting in a total of $12,000$ data shots. 

Scattered laser light fills the vacuum chamber during the experiment, resulting in plasma light emission images with high background noise at the laser frequency. The background noise level is hundreds of times higher for shots that include prompt laser light than for those that do not, making it easy to identify shots where the laser arrival time fell within the camera exposure window. 

Due to the high and fluctuating noise level, distinguishing the plasma signal from the background noise is somewhat challenging. A dynamic statistical threshold is employed to determine whether any particular image includes a signal from plasma emission light. A region of interest where the plasma light may appear is identified as the ``plasma region'' and the rest of the image is tagged as the ``background region''. We work out the $99.7\%$ error bound for a population mean (EBM), that is, the probability of the average of a random sample falling below $\overline{x}_{{\rm max}, 99.7}$ is $99.7\%$ based on
\begin{equation}
\overline{x}_{{\rm max}, 99.7}= \overline{x}+ z_{99.7} \frac{\sigma}{\sqrt{n}},
\label{eq:Zscore}
\end{equation}
where $z_{99.7}= 2.778$ is the $Z$ score for $99.7\%$ probability, $\overline{x}$ is the average light count value of the ``background region'', $\sigma$ is the standard deviation of the ``background region'', and $n$ is the sample size of the ``plasma region''. 
When the average value of the ``plasma region'' exceeds the threshold defined in Eq.~\ref{eq:Zscore}, it is considered a positive signal of plasma light emission. This dynamic statistical threshold effectively adapts to the varying background intensities and is more efficient than a fixed intensity threshold.

\begin{figure}
\includegraphics[width=1\columnwidth]{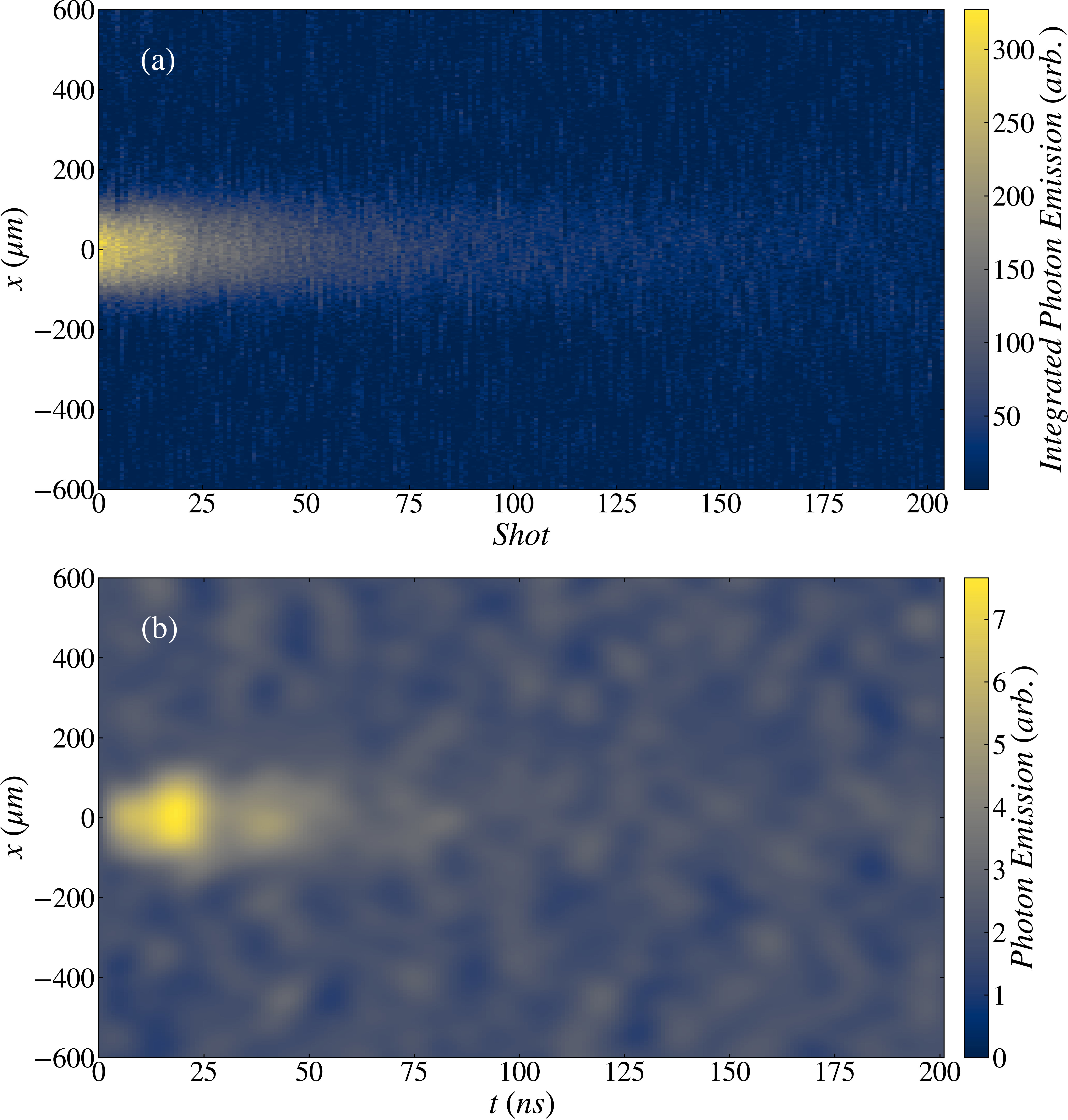}
\caption{\label{fig:sort} Each column in (a) represents a single image (shot) containing plasma light but without a laser signal. The images have been summed over the laser/plasma axial dimension. The columns are sorted and aligned based on their total intensity within the region of interest. Each image integrates from $t= t_{start}$ to $t_{EndEmission}$. In the shot with the highest intensity (leftmost column), $t_{start} \sim t_{laser}$; whereas in the rightmost column, $t_{start} \sim t_{EndEmission}$. (b) shows the result after a Gaussian (low-pass) filter is applied to (a) and each column is subtracted from the next column, yielding time-resolved (spatially integrated) photon emission profile with the temporal resolution $\sim 1$ ns.}
\end{figure}

By Eq.~\ref{eq:dt}, we determined that the decay time scale of the detectable signal for our laser-ionized Helium plasma light emission is $194 \pm 14 \,{\rm ns}$. We identified $204$ shots that exhibited plasma light emission without laser light based on Eq.~\ref{eq:Zscore}. Each image is rotated to ensure the plasma light filament appears completely horizontal. Since the plasma filament is homogeneous over the majority of its length, we took the projection of these images along the laser propagation axis and plotted them as a waterfall plot in Fig~\ref{fig:sort}(a), sorted by total intensity of the region of interest. Each column in the waterfall plot represents one plasma-light-only shot. Because the camera exposure window starts at $t_{\rm start}$, which is after the arrival of the laser pulse, $t_{\rm laser}$, the plasma light in each shot is integrated from $t_{\rm start}$ to the end of the detectable light emission, $t_{\rm endEmission}$. Since the decay time scale of the plasma light is orders of magnitude shorter than the characteristic camera jitter time scale, the $204$ plasma-light-only shots can be treated as being roughly uniformly distributed in time within the $194 \,{\rm ns}$ plasma light emission window, yielding an effective time resolution of approximately $1 \,{\rm ns}$. 

Every column in Fig.~\ref{fig:sort} corresponds to an integrated plasma light signal from $t_{\rm start}$ to $t_{\rm endEmission}$, (where $t_{\rm start}$ shifts $\sim 1 \,{\rm ns}$ further away from $t_{\rm laser}$ from one column to the next), thus subtracting each column from its neighboring column yields a time-resolved plasma light emission plot with a resolution of $\sim 1 \,{\rm ns}$. We apply a Gaussian noise filter to the waterfall plot in Fig.~\ref{fig:sort} (a), then subtract one column from the next to yield the time-resolved plasma light emission pattern shown in Fig.~\ref{fig:sort} (b). The results of this analysis demonstrate the ability to achieve $\mathcal{O}(1 \,{\rm ns})$ time-resolved imagery of the plasma light using a cost-effective GigE CMOS camera.


Finally, The camera records the light emission function projected along the camera axis, the radial emission function is retrieved with Abel inversion~\cite{hickstein2019}, as shown in Fig.~\ref{fig:4plots} (b).

\section{\label{sec:res}Results and Discussion}

\begin{figure}
\includegraphics[width=1\columnwidth]{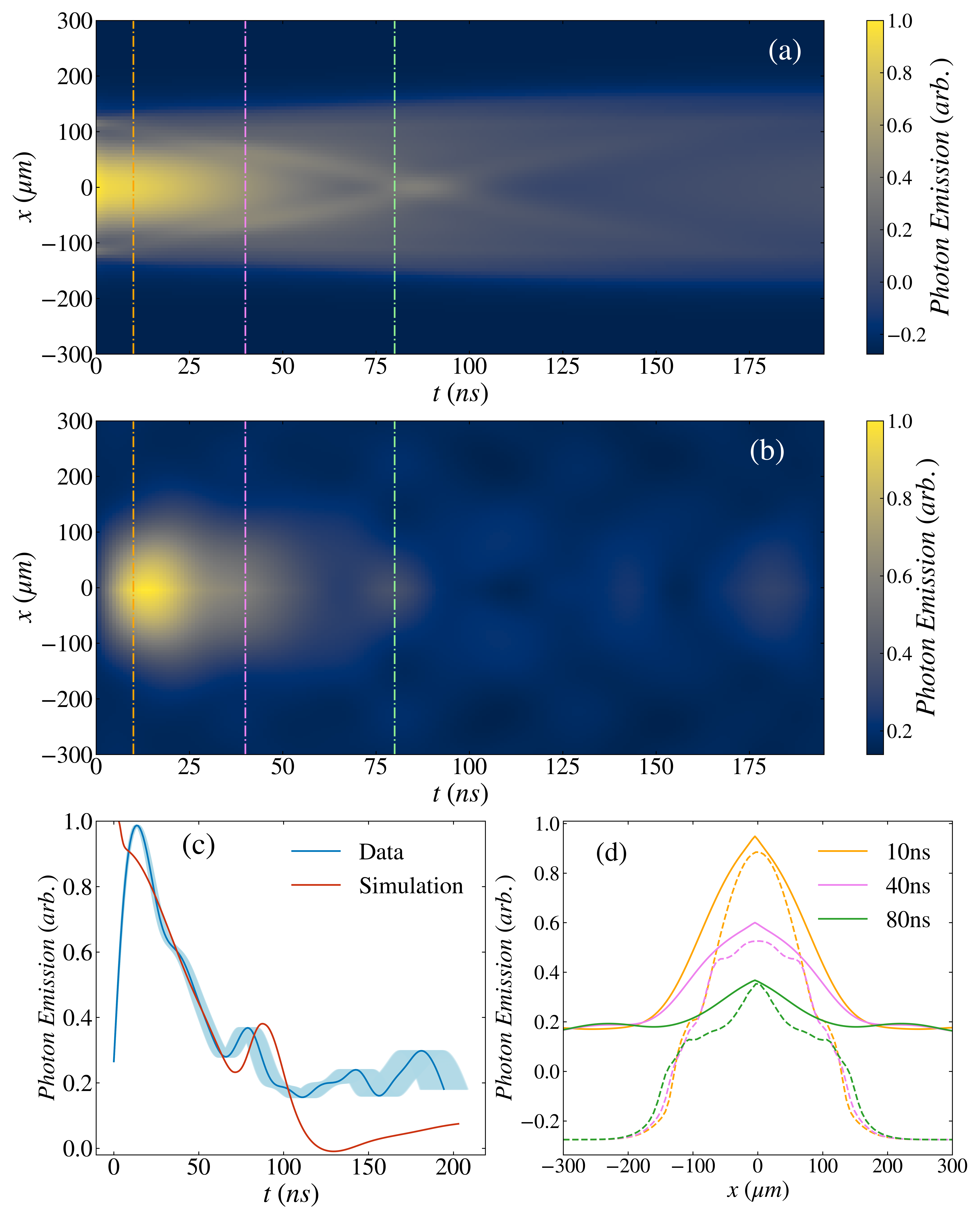}
\caption{\label{fig:4plots}Light emission from the plasma as a function of radius and time from (a) simulation and (b) measurement. Measurements were taken at $z=1$ m from the axicon. (c) shows the temporal evolution at $r=0$ with simulation in red and experimental data in blue. The shaded region is the statistical temporal measurement error. (d) shows radial lineouts at 10, 40, and 80ns, solid lines are experiment and dashed lines simulation.}
\end{figure}

We compare the results of our simulation work and our experimental data analysis in Fig.~\ref{fig:4plots}, and observe good agreement overall. Figure~\ref{fig:4plots} (a) and (b) show the time evolution of the simulated and experimentally observed plasma light emission along the $x$ axis ($y=0$) at an arbitrary $z$ location within the homogeneous middle region of the plasma filament, respectively. The simulations used to generate Fig.~\ref{fig:4plots} (a) are described in section~\ref{sec:modeling}, and the data analysis methods lead to Fig.~\ref{fig:4plots} (b) is described in Section \ref{sec:DataPro}. Figure~\ref{fig:4plots} (a) and (b) are normalized for comparison.
 
Figure~\ref{fig:4plots} (c) shows the photon emission density from Fig.~\ref{fig:4plots} (a) and (b) at the center location ($x=0, y=0$) as a function of time. We see good agreement, including the approximate time of the second peak in brightness between 75 and 100 nanoseconds. There is a slight discrepancy in the first $10 \ {\rm ns}$, where the data shows a sharp rise from zero, and the simulation immediately yields a high photon emission density. This is due to the fact that the model we used in our simulation does not take into account the atomic decay lifetime of the excited atoms, which leads to a slight delay between the formation of the plasma and the onset of photon emission. The lifetime of the $4p$ state is 3.975 ns and the $3d$ state is 15.696 ns~\cite{Theodosiou1987}, which qualitatively matches the delayed peak observed at 13.5 ns.

Figure~\ref{fig:4plots} (d) compares the full photon emission density profile along $x$ at a few different times, $10$, $40$, and $80 \ {\rm ns}$. We again see generally good agreement, though the data does not seem to capture the $\sim 400\,{\rm \mu m}$-wide diffuse region of light at later times. This is likely an artifact of our limited signal-to-noise ratio in the experimental data, and the noise level can be observed in Fig~\ref{fig:4plots} (c) as well after $\sim100$ ns.

\begin{figure}
\includegraphics[width=1\columnwidth]{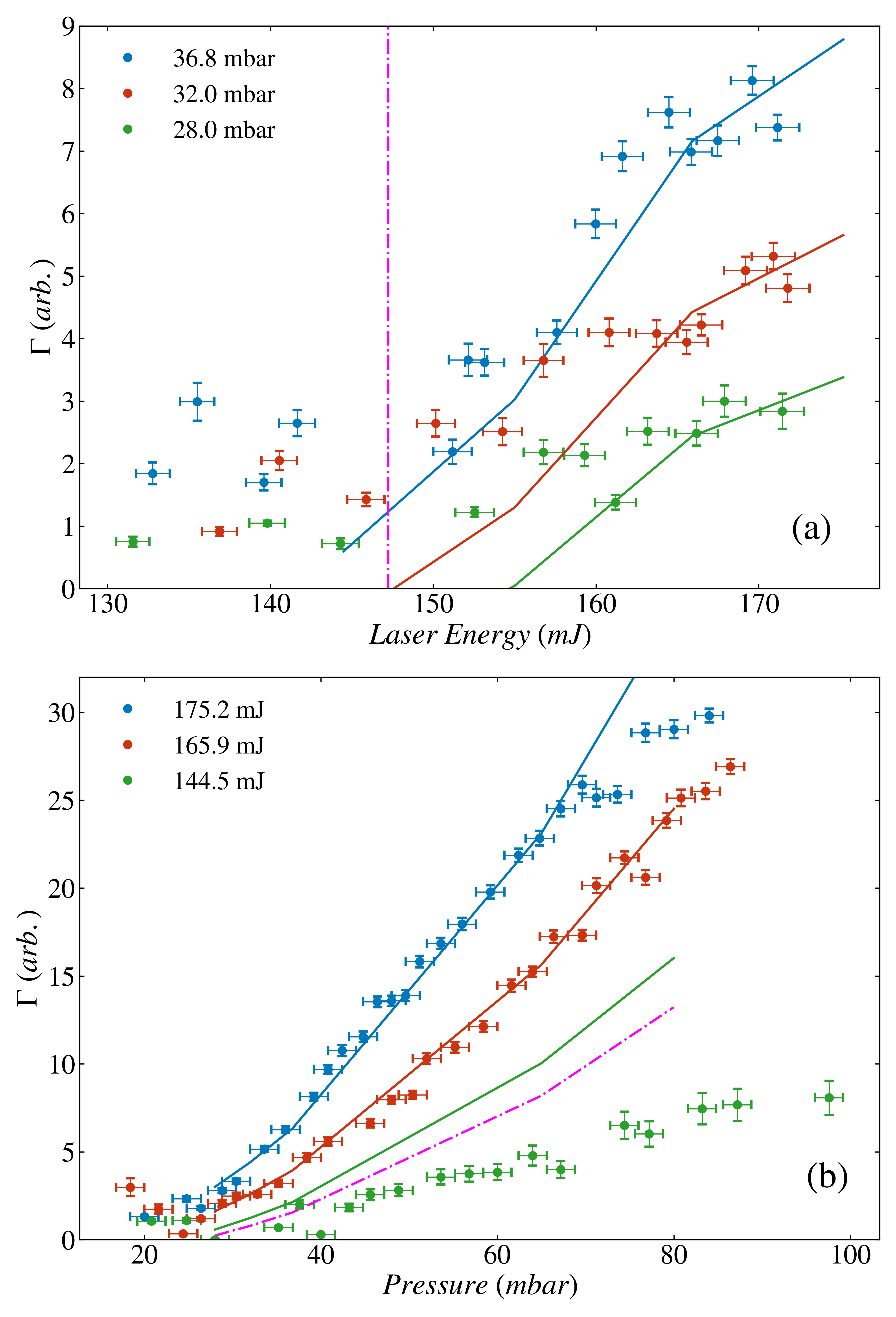}
\caption{\label{fig:Scan} 
{\em Time-integrated} peak photon emission as a function of (a) laser energy and (b) gas pressure. Each data point represents the average result of 100 measurements. The solid lines show the calculated quantity from Eq.~\ref{eq:ana}, where $T_e$ is simulated using PIC simulations and $n_0$ is assumed to correspond to a fully ionized gas at the given pressure. The purple dot-dashed line in (a) shows the laser energy at which $99\%$ ionization occurs, approximately $147\,{\rm mJ}$. The purple dot-dashed line in (b) shows the curve corresponding to calculation at which $99\%$ ionization occurs, below which the gas is not expected to be fully ionized. The analytic formula agrees well with the data when the plasma is fully ionized.}
\end{figure}

Lastly, Figure~\ref{fig:Scan} (a) and (b) show the experimental results of a laser energy scan and a gas pressure scan, respectively. The data points of each plot correspond to the peak photon emission observed in the {\em time-integrated} ({\it i.e.} the entire plasma light lifetime falls within the camera integration time), spatially resolved data (average of 100 shots). The lines correspond to an analytic prediction of the temporally integrated peak photon emission, $\Gamma$, using Eq.~\ref{eq:ana}. Despite the complexity of the plasma light emission process, the experimental data shows remarkable agreement with our simple prediction of the peak photon density scaling as a function of the initial plasma parameters, $n_0$ and $T_0$.

\section{\label{sec:con}Conclusions}

We have demonstrated an experimental comprehension of the decay process of the plasma light emitted from a thin, laser-ionized Helium plasma source, supported by numerical simulations. We used a three-step model to model the photon emission pattern, which showed that electron-neutral collisional excitation dominates over electron-ion recombination. The photon emission pattern was measured in the experiment from a $6.9\times 10^{17} \ {\rm cm^{-3}}$, $\sim 30 \rm{\mu m}$ wide plasma and agrees well with the simulation. We constructed an analytic model to predict the scaling of the time-integrated, on-axis light emission with the initial plasma density and temperature according to Eq.~\ref{eq:ana} and we saw the expected scaling in the experiment.

We also presented a novel statistical approach for measuring the temporal evolution of plasma light with nanosecond resolution using a cost-effective GigE CMOS machine vision camera. By leveraging the prompt scattered laser light as a fiducial signal, we quantified the exposure timing jitter distribution of the camera and measured the observable plasma light lifetime at $194 \pm 14 \,{\rm ns}$. Furthermore, by sorting and applying image processing techniques, we reconstructed the time-resolved plasma light emission profile with nanosecond-level precision. 


This statistical method 
is generalizable and offers new diagnostic possibilities in PWFA experiments and related applications where cost-effective diagnostics are a necessity. To prove effective, there must exist an appropriate temporal fiducial signal (e.g. prompt laser light) and the the detector’s timing jitter distribution must exhibit statistical repeatability. In PWFA experiments, these two requirements are often met due to the presence of a laser and an electron beam, which can serve as reliable temporal fiducial signals. This technique allows commonly used sensors to achieve measurements previously exclusive to expensive gated sensors ({\it e.g.} gated cameras or gated spectrometers) on a statistical basis, giving inexpensive GigE CMOS cameras access to nanosecond dynamics. For example, this technique could be applied to Stark broadening of emission lines in the plasma glow~\cite{GRrKM1962} detected by a spectrometer; a CMOS camera equipped with a wide-angle lens can capture the formation of laser-ionized meter-scale plasma. Some potential applications in PWFA include imagining the dissipation of the plasma wake~\cite{wan2022}, investigating the plasma heating mechanism by an electron beam for laser-electron alignment purposes~\cite{Scherkl2022}, and measuring the recovery time of a plasma-wakefield accelerator~\cite{darcy2022}.

Finally, we come to two more significant conclusions regarding the work presented here. First, our results enhance our confidence in the validity of the plasma formation models. In particular, we have confirmed that the relatively inexpensive SSF simulation is able to accurately predict the laser ionization process, making it a valuable tool in the iterative plasma source design process. Second, our advances in understanding of the readily accessible plasma light emission have enhanced its utility as a diagnostic tool. Numerous studies have shown that the electron temperature of the PWFA-like plasma source impacts a wide range of applications~\cite{Diederichs, ma2020, miao2022}, and the methods described in our work can make it possible to design and diagnose experiments that aim to study such effects with significantly greater confidence. 



\begin{acknowledgments}
This material is based upon work supported by the U.S. Department of Energy, Office of Science, Office of High Energy Physics under Award No. DE-SC0017906 and the National Science Foundation under Grant Number PHY-2047083.
\end{acknowledgments}

\appendix

\section{Excitation and recombination model}

Detectable visible/NIR photons are emitted by transitions in Helium between the $n=3$ and $n=2$ energy levels. Among these transitions, the $3d-2p$ transition exhibits the highest rate and the most persistent atomic spectral line of neutral Helium~\cite{Theodosiou1987}. Consequently, our focus is on processes that populate the $3d$ state. To populate the $3d$ state, electrons can transition from $p$ or $f$ states. Although the $4f-3d$ transition has a higher transition than the $4p-3d$ transition, populating the $f$ states requires a higher excitation energy, which corresponds to a significantly smaller collisional cross-section. Therefore, in a collisional excitation scenario, the most probable transition from the ground state that will yield a visible/NIR photon is the sequence $1s-4p-3d-2p$.

On the other hand, in plasma electron-ion recombination, it takes less energy to capture a free electron into a highly excited state~\cite{Rqbben1963}. As a result, most recombination events result in highly excited states, thereby easily populating the $4f$ state. Therefore, in a recombination and de-excitation process, the most probable transition leading to a visible/NIR photon is $4f-3d-2p$.

\section{Temporally integrated peak photon emission density}

\begin{figure}
\includegraphics[width=1\columnwidth]{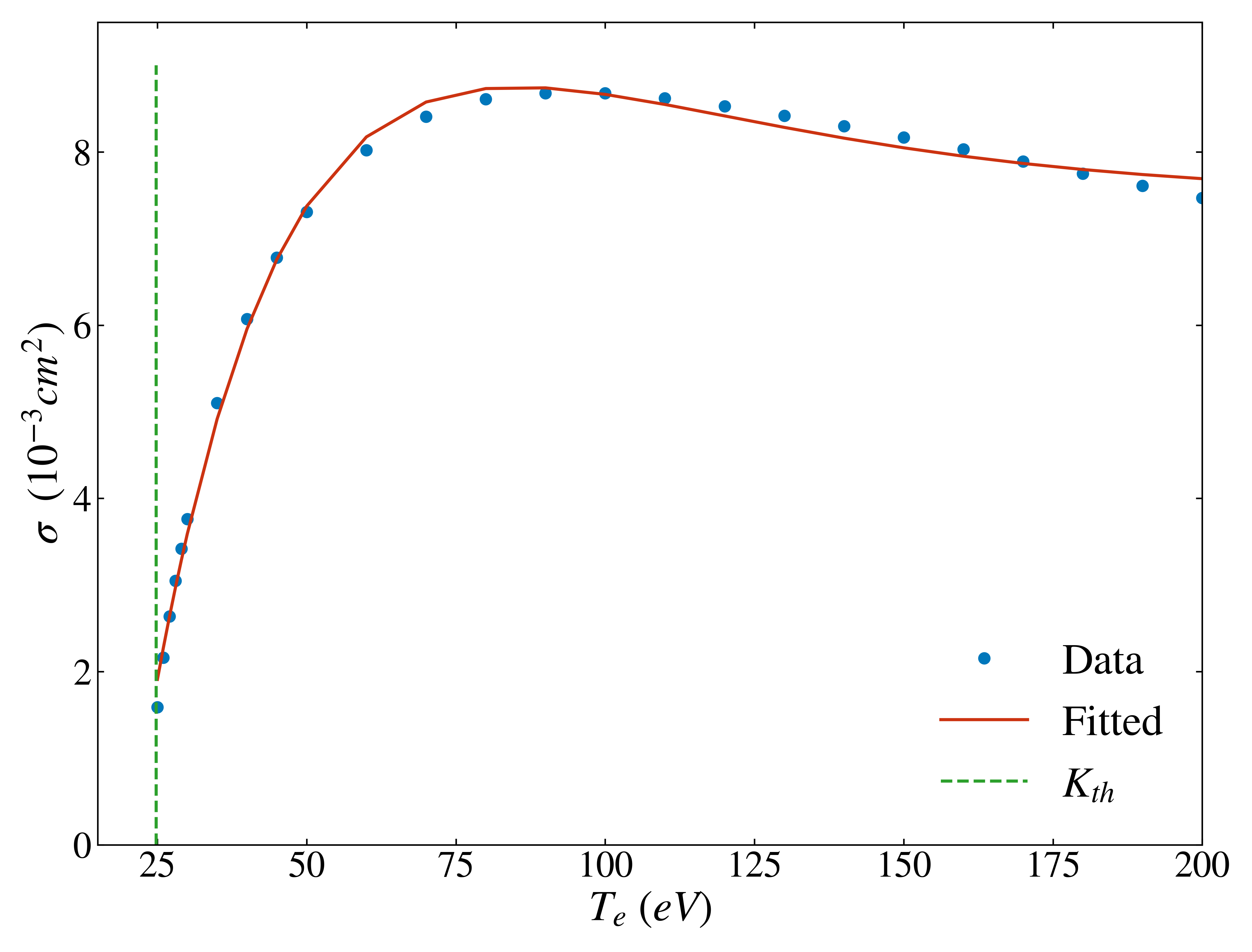}
\caption{\label{fig:Xsec} The electron-impact excitation cross-section of the $1s$ state to the $4p$ state, $\sigma_{1s-4p}$. The blue data points are taken from Ref.~\onlinecite{stone2002} and the red curve corresponds to a fit using Eq.~\ref{eq:fit}. The green dashed line denotes the excitation threshold energy, $K_{th}$.}
\end{figure}

\begin{figure}
\includegraphics[width=1\columnwidth]{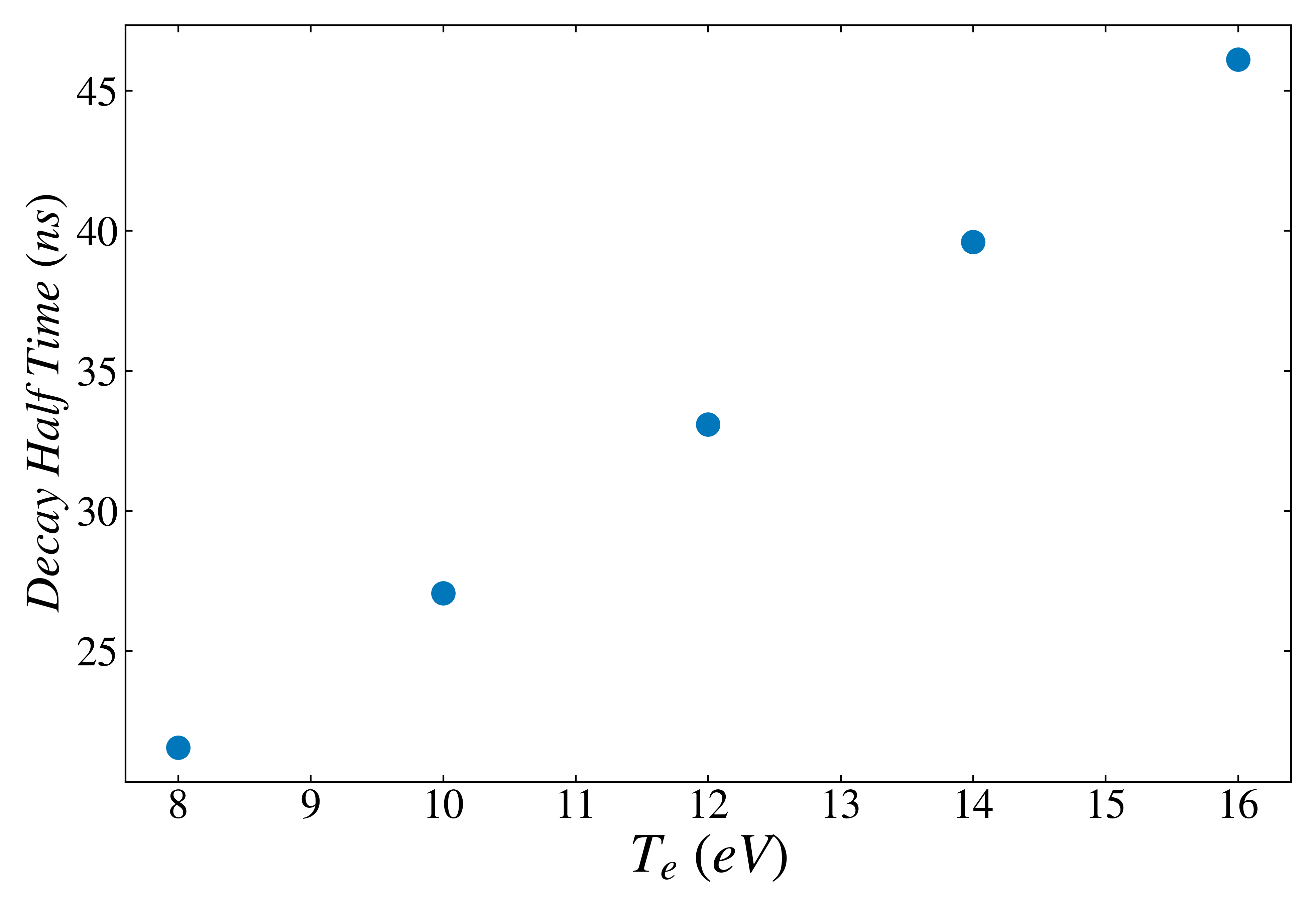}
\caption{\label{fig:HalfTime} The peak photon density decay half-life (at $x=0, y=0$) for various initial average electron temperatures for a thin, laser ionized PWFA plasma with density $n_e= 8e17 \ {\rm cm^{-3}}$.}
\end{figure}

The analytic formalism for the temporally integrated peak photon emission is derived by integrating $n_e<\nu_{en}>$ from $t=0$ to $t=\infty$ with a few key approximations and assumptions. Recall $<\nu_{en}>$ is given by Eq.~\ref{eq:NKiMaxwellian}. We note that the peak photon emission occurs at $\vec{r} \simeq 0$ and for simplification, we drop the notation $\vec{r} \simeq 0$. To calculate the integral in Eq.~\ref{eq:NKiMaxwellian}, we first plot the electron-impact excitation cross-section from Ref.~\onlinecite{stone2002} in Fig.~\ref{fig:Xsec}, and fit it with the following function:
\begin{eqnarray}
\sigma (K)_{1s-4p}= A(K-\kappa-K_{th})\exp({-\frac{K}{K_w}})+\psi, 
\label{eq:fit}
\end{eqnarray}
where $A= 2.1 \times 10^{-4} \ {\rm cm^{2}}, \kappa= 34.2 \ {\rm eV}, K_w= 26 \ {\rm eV}, \psi=7.5 \times 10^{-3} \ {\rm cm^{2}}$ are the free parameters found by the fit. The integral in Eq.~\ref{eq:NKiMaxwellian} becomes
\begin{eqnarray}
I_{exc} \approx A \int_{K_{th}}^{\infty} &&\exp({-\frac{K}{k_B T_e(t)}}) \nonumber\\
&&((K-\kappa)\exp({-\frac{K}{K_w}})+\psi) K dK.
\label{eq:integral}
\end{eqnarray}
Because the electron temperature during the plasma glowing process (as shown in Fig.~\ref{fig:expand} (b)) is orders of magnitude less than the initial temperature, we make the following approximations, $K_{th}>>k_BT_e$, $C>>k_BT_e$, and $B>>k_BT_e$. The integral now becomes
\begin{eqnarray}
I_{exc} \approx A k_BT_e(t) \exp({-\frac{K_{th}}{k_B T_e(t)}})(\psi K_{th}- \sigma_{1s-4p}),
\label{eq:integralreduced}
\end{eqnarray}
where $\sigma_{1s-4p}= \frac{4}{K_w^2} \exp{(\frac{K_{th}}{K_w}) K_{th}^4}$. Now, we plug the result of the integral in Eq.~\ref{eq:integralreduced} back into Eq.~\ref{eq:NKiMaxwellian}, 
\begin{eqnarray}
n_e(t)<\upsilon_{en}(t)> \approx && A' n_e(t) n_n(t) (k_B T_e(t))^{-3/2} \nonumber\\
&&k_BT_e(t) \exp({-\frac{K_{th}}{k_b T_e(t)}}),
\label{eq:density rate}
\end{eqnarray}
where $A'$ is a collective constant. To calculate the temporally integrated photon emission, we integrate Eq.~\ref{eq:density rate} from $t=0$ to $t=\infty$,
\begin{eqnarray}
&&\int_{0}^{\infty} n_e(t)<\upsilon_{en}(t)> dt \nonumber\\
&& \approx \int_{0}^{\infty} A' n_e(t) n_n(t) (k_B T(t))^{-1/2} \exp({-\frac{K_{th}}{k_B T(t)}}) dt.
\label{eq:timeInt}
\end{eqnarray}
To continue the calculation, a few assumptions are made: $n_n(t)= n_0$, $k_BT_e(t)\approx k_B T_{e0} e^{-t/\tau_d}$, $n_e(t)\approx n_0 e^{-t/\tau_d}$, where $\tau_d$ is the decay lifetime and $T_{e0}$ is the plasma electron temperature before the decay. The first assumption is justified because the neutral particles fill in the plasma column much faster than the full expansion time scale, leading to a quasi-homogenous background of neutral gas. The second and third assumptions are justified by the lineout in Fig.~\ref{fig:expand}(b). So Eq.~\ref{eq:timeInt} now becomes
\begin{eqnarray}
\int_{0}^{\infty} &&
A' n_0^2 e^{-t/\tau_d} (k_B T_{e0} e^{-t/\tau_d})^{-1/2} 
\exp\left(
{-\frac{K_{th}}{k_B T_{e0} e^{-t/\tau_d}}}
\right) dt \nonumber\\
\approx && A' n_0^2 (k_B T_{e0})^{-1/2}2\tau_d
\nonumber \\
&&\left(e^{-\frac{K_{th}}{k_BT_{e0}}}- \frac{K_{th}\sqrt{\pi} \text{Erfc}(\sqrt{\frac{K{th}}{k_b T_{e0}}})}{\sqrt{K_{th} k_B T_{e0}}}\right), 
\label{eq:timeInt2}
\end{eqnarray}
where Erfc is the complementary error function, and when $k_B T_{e0}<< K_{th}$, $\text{Erfc}(\sqrt{\frac{K{th}}{k_b T_{e0}}}) \approx0$. This simplifies Eq.~\ref{eq:timeInt2} to
\begin{eqnarray}
&&\int_{0}^{\infty} n_e(t)<\upsilon_{en}(t)> dt \nonumber\\
\approx && A' n_0^2 (k_B T_{e0})^{-1/2}2\tau_de^{-\frac{K_{th}}{k_BT_{e0}}}, 
\label{eq:timeInt3}
\end{eqnarray}
We empirically find from our simulations that the decay lifetime $\tau_d$ is linearly proportional to the initial kinetic energy as shown in Fig.~\ref{fig:HalfTime}. So Eq.~\ref{eq:timeInt3} becomes
\begin{eqnarray}
\int_{0}^{\infty} n_e(t)<\upsilon_{en}(t)> dt \approx 2A' n_0^2 (k_B T_{e0})^{1/2} e^{-\frac{K_{th}}{k_BT_{e0}}}.
\label{eq:final}
\end{eqnarray}
Finally, we note that the observed plasma light arises almost entirely from collisions occurring after the initial rapid plasma expansion, so we plug in the temperature after the initial rapid expansion $T_{e0}= RT_0$ where we empirically find $R= 0.09$ from our simulations. Thus, we finally arrive at the following expression of proportionality for the time-integrated peak light emission $n_{\gamma,{\rm peak}}$:
\begin{eqnarray}
\Gamma \propto
n_0^2 \sqrt{R k_B T_0} \exp\left({-\frac{K_{th}}{Rk_BT_0}}\right).
\label{eq:finalfinal}
\end{eqnarray}

\bibliography{aipsamp}

\end{document}